\documentclass[conference]{IEEEtran}
\IEEEoverridecommandlockouts

\usepackage{cite}
\usepackage{amsmath,amssymb,amsfonts}
\usepackage{bm}
\usepackage{algorithmic}
\usepackage{graphicx}
\usepackage{textcomp}
\usepackage{xcolor}
\usepackage{booktabs}
\usepackage{multirow}

\def\BibTeX{{\rm B\kern-.05em{\sc i\kern-.025em b}\kern-.08em
    T\kern-.1667em\lower.7ex\hbox{E}\kern-.125emX}}

\begin{document}


\setlength{\abovedisplayskip}{3pt}
\setlength{\belowdisplayskip}{3pt}
\setlength{\abovedisplayshortskip}{2pt}
\setlength{\belowdisplayshortskip}{2pt}
\setlength{\textfloatsep}{6pt}
\setlength{\floatsep}{5pt}
\setlength{\intextsep}{5pt}


\title{Ray-Tracing-Parameter-Based Lightweight Hybrid Precoding via Cascade XGBoost for Wideband Massive MIMO-OFDM Systems
\vspace{-0.15in}
}

\author{
\IEEEauthorblockN{
Po-Heng Chou$^{1,3}$,
Wei-Chen Lin$^{2}$,
Wan-Jen Huang$^{2}$,
Walid Saad$^{3}$,
and Ronald Y. Chang$^{1}$
\thanks{This work was supported in part by the National Science and Technology Council (NSTC) of Taiwan under Grant 113-2926-I-001-502-G. Prof. Walid Saad was supported by the U.S. National Science Foundation (NSF) under Grant CNS-2114267.}
}
\IEEEauthorblockA{
$^{1}$Research Center for Information Technology Innovation (CITI),
Academia Sinica (AS), Taipei 11529, Taiwan\\
$^{2}$Institute of Communication Engineering (ICE),
National Sun Yat-sen University (NSYSU), Kaohsiung 80424, Taiwan\\
$^{3}$Bradley Department of Electrical and Computer Engineering (ECE),
Virginia Tech (VT), Alexandria, VA 22305, USA\\
E-mails: d00942015@ntu.edu.tw, wilson20020501@gmail.com,
wjhuang@faculty.nsysu.edu.tw,\\ walids@vt.edu, rchang@citi.sinica.edu.tw
\vspace{-0.25in}
}
}

\maketitle


\begin{abstract}
Hybrid precoding reduces the radio-frequency (RF) hardware burden in wideband massive MIMO-OFDM systems, but conventional optimization and deep neural network (DNN)-based approaches can incur high online complexity. This paper proposes XGB-SVD, a lightweight hybrid precoding framework that maps compact ray-tracing path parameters directly to analog beamforming codeword indices using cascade eXtreme Gradient Boosting (XGBoost). The cascade captures sequential beam-selection dependencies and transmit--receive propagation correlation, while the digital precoder and combiner are obtained from the reduced-dimensional equivalent channel using singular value decomposition (SVD). Evaluations over five DeepMIMO scenarios spanning sub-6-GHz and millimeter-wave bands show that XGB-SVD consistently outperforms the considered DNN baselines and approaches simultaneous orthogonal matching pursuit (SOMP) in several environments. Cross-scenario results further indicate improved transfer from propagation-diverse training environments, while XGB-SVD provides lower online computational and training costs than the DNN baselines.
\end{abstract}



\section{Introduction}

Wideband massive multiple-input multiple-output orthogonal frequency-division multiplexing (MIMO-OFDM) is expected to play an important role in future sixth-generation (6G) wireless systems operating across both sub-6-GHz and millimeter-wave (mmWave) spectrum~\cite{liu2025itu}. Large antenna arrays provide substantial beamforming and spatial multiplexing gains, but fully digital architectures require one radio-frequency (RF) chain per antenna element, resulting in high hardware cost and power consumption.

Hybrid analog-digital precoding reduces the RF-chain burden by combining digital processing with an analog phase-shifter network. However, the constant-modulus constraint of the analog stage and the use of a common analog precoder across OFDM subcarriers make wideband hybrid precoding a challenging non-convex optimization problem.

Conventional hybrid precoding methods based on orthogonal matching pursuit (OMP), simultaneous orthogonal matching pursuit (SOMP), and alternating minimization can achieve high spectral efficiency ~\cite{ayach2014spatial,lee2015hybrid,yu2016alternating,sohrabi2017hybrid}. Nevertheless, repeated matrix operations and iterative optimization can incur substantial online complexity as the system dimension increases.

Learning-based hybrid precoding shifts part of this computational burden
from online optimization to offline training. Deep neural networks
(DNNs) and convolutional neural networks (CNNs) have been used to learn
the mapping from channel-domain representations to analog and digital
beamformers~\cite{elbir2019cnn,li2019deep,elbir2021family,mir2019machine}.
More recent studies have investigated low-complexity neural architectures,
cross-band channel reconstruction, and feedback-efficient hybrid precoding
~\cite{liu2025lowcomplexity,wei2026crossband,chou2026reducedfeedback}.
However, these approaches still commonly rely on high-dimensional channel
representations, explicit CSI reconstruction, neural-network inference,
or model-based feedback processing.

Physical propagation parameters provide a compact alternative to
full channel-matrix representations. Sensing-assisted communication and parametric channel estimation can
provide angles of arrival (AoAs), angles of departure (AoDs), path gains,
phases, and delays~\cite{xu2024little,wang2025sensing}, while recent integrated sensing and communication (ISAC) studies
further connect such parameters to channel modeling and hybrid
precoding~\cite{wang2025widebandisac}. These parameters can
therefore support analog beam selection without reconstructing the full
frequency-domain channel during learning-based inference.

Motivated by this observation, this paper develops a lightweight
hybrid precoding framework based on eXtreme Gradient Boosting
(XGBoost)~\cite{chen2016xgboost}. The proposed method directly maps
compact ray-tracing path parameters to discrete analog beamforming
codeword indices, while a cascade XGBoost architecture captures the
sequential dependence among beam selections and the correlation between
transmit and receive propagation directions. After analog beam
selection, the digital precoder and combiner are obtained analytically
from the reduced-dimensional equivalent channel using singular value
decomposition (SVD). Thus, the proposed framework combines data-driven analog beam
selection with model-based digital processing after RF-domain
dimensionality reduction.

The main contributions of this work are summarized as follows:

\begin{itemize}\setlength{\itemsep}{0pt}

\item We develop a physical-parameter-driven hybrid precoding framework that predicts analog beamforming codewords directly from compact ray-tracing path parameters, avoiding explicit full-dimensional channel reconstruction during the machine-learning-based analog beam prediction stage.

\item We propose a cascade XGBoost architecture that exploits both the sequential dependence of SOMP beam selection and the physical correlation between transmit and receive propagation directions.

\item We combine learned analog beam selection with reduced-dimensional SVD-based digital processing, avoiding direct regression of complex-valued digital beamforming coefficients and providing a favorable performance-complexity trade-off.

\item We evaluate the proposed framework over five DeepMIMO ray-tracing environments spanning sub-6-GHz and mmWave operation. Same-scenario, multi-stream, and cross-scenario experiments are conducted to characterize spectral-efficiency performance, computational efficiency, and transfer behavior across propagation environments.

\end{itemize}


\section{System Model and Problem Formulation}

\subsection{Wideband Hybrid Precoding Model}

We consider a wideband MIMO-OFDM system with
a hybrid precoder at the transmitter and a hybrid combiner at the receiver.
The transmitter is equipped with $N_t$ antennas and $N_t^{\mathrm{RF}}$ RF chains,
while the receiver is equipped with $N_r$ antennas and $N_r^{\mathrm{RF}}$ RF
chains. The transmitted OFDM waveform is composed of $K$ subcarriers, and $N_s$
spatial streams are conveyed over each subcarrier, where
$N_s \leq N_t^{\mathrm{RF}} \ll N_t$ and
$N_s \leq N_r^{\mathrm{RF}} \ll N_r$.

The  baseband signal transmitted over the $k$-th subcarrier $\bm{x}[k]\in\mathbb{C}^{N_t\times1}, k\in\{1,\ldots,K\}$ is
\begin{equation}
\bm{x}[k]
=
\sqrt{\frac{P_t}{N_s}}\,
\bm{F}_{\mathrm{RF}}
\bm{F}_{\mathrm{D}}[k]
\bm{s}[k],
\label{eq_tx}
\end{equation}
where $P_t$ is the total transmit power per subcarrier,
$\bm{s}[k]\in\mathbb{C}^{N_s\times1}$ is the symbol 
vector conveyed over the $k$-th subcarrier, satisfying
$\mathbb{E}\{\bm{s}[k]\bm{s}^{H}[k]\}=\bm{I}_{N_s}$.
The matrices  $\bm{F}_{\mathrm{RF}}
\in\mathbb{C}^{N_t\times N_t^{\mathrm{RF}}}$ and 
$\bm{F}_{\mathrm{D}}[k]
\in\mathbb{C}^{N_t^{\mathrm{RF}}\times N_s}$
are the analog precoder and the digital precoder, respectively. The analog precoder is composed of phase shifters; thus, the entries in $\bm{F}_{\mathrm{RF}}$
satisfy the constant-modulus constraint
$|[\bm{F}_{\mathrm{RF}}]_{m,n}|=1/\sqrt{N_t}$,
where $[\bm{F}_{\mathrm{RF}}]_{m,n}$ denotes its $(m,n)$th element.

The baseband received signal over the subcarrier $k$ is
\begin{equation}
\bm{y}[k]
=
\bm{H}[k]\bm{x}[k]
+
\bm{n}[k],
\label{eq_rx}
\end{equation}
where
$\bm{H}[k]\in\mathbb{C}^{N_r\times N_t}$
is the frequency response of the  MIMO channel, and
$\bm{n}[k]\sim
\mathcal{CN}(\bm{0},\sigma_n^2\bm{I}_{N_r})$
is the additive white Gaussian noise vector with noise variance
$\sigma_n^2$.
The receiver first applies an analog combiner, followed by subcarrier-dependent digital combiners.
The detected symbol vector is
\begin{equation}
\hat{\bm{s}}[k]
=
\bm{W}_{\mathrm{D}}^{H}[k]
\bm{W}_{\mathrm{RF}}^{H}
\bm{y}[k],
\label{eq_combined}
\end{equation}
where $\bm{W}_{\mathrm{RF}}
\in\mathbb{C}^{N_r\times N_r^{\mathrm{RF}}}$ and 
$\bm{W}_{\mathrm{D}}[k]
\in\mathbb{C}^{N_r^{\mathrm{RF}}\times N_s}$ are the analog combiner and the digital combiner applied over the $k$-th subcarrier. Similarly, the analog combiner  satisfies
$|[\bm{W}_{\mathrm{RF}}]_{m,n}|=1/\sqrt{N_r}$.


\begin{figure*}[t]
\centering
\includegraphics[width=\textwidth]{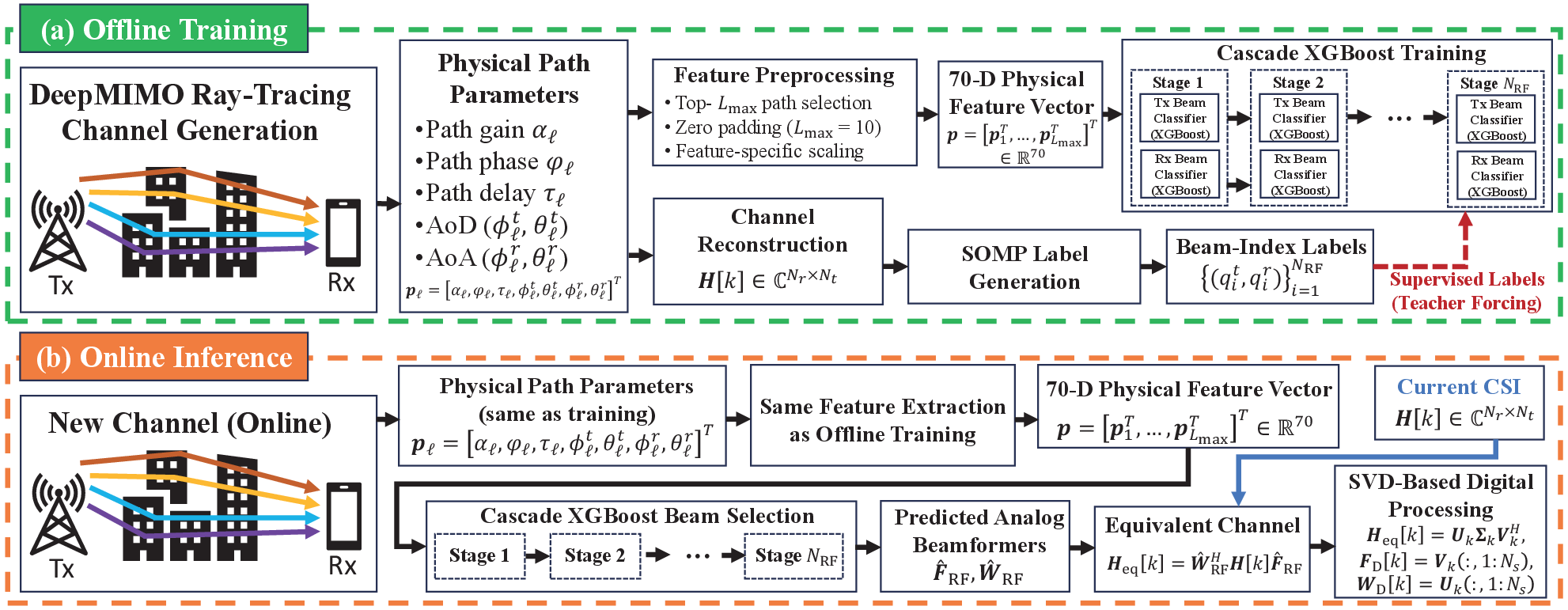}
\caption{Proposed XGB-SVD framework with (a) offline SOMP-based label generation and (b) online Cascade XGBoost beam selection followed by SVD-based digital processing.}
\label{fig_framework}
\vspace{-0.2in}
\end{figure*}
\vspace{-0.2in}
\subsection{Ray-Tracing Channel Representation}

In the ray-tracing model, a MIMO channel is represented through a finite set of
physically interpretable propagation paths.
At subcarrier $k$, the channel is modeled as
\begin{align}
\bm{H}[k]
=
\sqrt{\frac{N_tN_r}{L}}
\sum_{\ell=1}^{L}
\alpha_{\ell}
e^{j\varphi_{\ell}}
e^{-j2\pi f_k\tau_{\ell}}
\bm{a}_{r}
(\phi_{\ell}^{r},\theta_{\ell}^{r})
\bm{a}_{t}^{H}
(\phi_{\ell}^{t},\theta_{\ell}^{t}),
\label{eq_channel}
\end{align}
where $L$ denotes the number of propagation paths, $\ell$ is the
path index, $\alpha_{\ell}$ is the path amplitude,
$\varphi_{\ell}$ is the path phase, $\tau_{\ell}$ is the propagation
delay, $f_k$ denotes the frequency offset of subcarrier $k$ relative to the carrier frequency, and
$j=\sqrt{-1}$. The angles
$\phi_{\ell}^{t}$ and $\phi_{\ell}^{r}$ denote the transmit and receive
azimuth angles, respectively, while
$\theta_{\ell}^{t}$ and $\theta_{\ell}^{r}$ denote the corresponding
elevation angles.
The vectors
$\bm{a}_{t}(\phi,\theta)\in\mathbb{C}^{N_t\times1}$ and
$\bm{a}_{r}(\phi,\theta)\in\mathbb{C}^{N_r\times1}$
denote the three-dimensional beam-steering vectors for the uniform planar arrays (UPAs) at the transmitter and receiver, respectively.

In most existing work, the hybrid precoder is optimized based on the information of the channel matrix. To reduce the learning model size and the corresponding computational complexity, the channel information adopted in our model is replaced with a vector of the  path-domain features, which is defined by
$\bm{p}_{\ell}=
[\alpha_{\ell},\varphi_{\ell},\tau_{\ell},
\phi_{\ell}^{t},\theta_{\ell}^{t},
\phi_{\ell}^{r},\theta_{\ell}^{r}]^{T},$
for each propagation path.
The input of the proposed model is a fixed-dimensional feature representation obtained by retaining at
most the $L_{\max}$ strongest paths and stacking them as
$\bm{p}=[\bm{p}_{1}^{T},\ldots,\bm{p}_{L_{\max}}^{T}]^{T}
\in\mathbb{R}^{7L_{\max}}$.
If $L>L_{\max}$, the $L_{\max}$ strongest paths are retained, whereas
zero padding is applied when $L<L_{\max}$. In this work, the length is set as $L_{\max}=10$. Notably, the length of the learning input is independent of the number of OFDM subcarriers and UPA dimensions.
\vspace{-0.2in}

\subsection{Problem Formulation of the Hybrid Precoding Optimization}

Define the composite precoding and combining matrices as
$\bm{F}[k]=\bm{F}_{\mathrm{RF}}\bm{F}_{\mathrm{D}}[k]$
and
$\bm{W}[k]=\bm{W}_{\mathrm{RF}}\bm{W}_{\mathrm{D}}[k]$,
respectively. The achievable spectral efficiency is
\begin{align}
R
=&\frac{1}{K}\sum_{k=1}^{K}
\log_2\det\Big(
\bm{I}_{N_s}
+\frac{\rho}{N_s}
\bm{R}_{n}^{-1}[k]
\bm{W}^{H}[k]\bm{H}[k]\bm{F}[k]
\nonumber\\
&\qquad\qquad\quad\times
\bm{F}^{H}[k]\bm{H}^{H}[k]\bm{W}[k]
\Big),
\label{eq_rate}
\end{align}
where $\rho=P_t/\sigma_n^2$ denotes the
transmit signal-to-noise ratio (SNR).
The matrix
$\bm{R}_n[k]=\bm{W}^{H}[k]\bm{W}[k]$
denotes the normalized post-combining
noise covariance matrix, while the
actual noise covariance matrix is
$\sigma_n^2\bm{R}_n[k]$.
For a fixed number of spatial streams with
equal power allocation, the reference fully
digital precoder
$\bm{F}_{\mathrm{opt}}[k]\in
\mathbb{C}^{N_t\times N_s}$
and combiner
$\bm{W}_{\mathrm{opt}}[k]\in
\mathbb{C}^{N_r\times N_s}$
are constructed from the dominant $N_s$
right and left singular vectors of
$\bm{H}[k]$, respectively.


The optimization of the analog precoder and digital precoders in terms of the maximal achievable rate can be expressed as follows ~\cite{lee2015hybrid}, 
\begin{subequations}
\label{precoding_problem}
\begin{align}
\min_{\bm{F}_{\mathrm{RF}},
\{\bm{F}_{\mathrm{D}}[k]\}_{k=1}^{K}}
\quad&
\sum_{k=1}^{K}
\left\|
\bm{F}_{\mathrm{opt}}[k]
-
\bm{F}_{\mathrm{RF}}
\bm{F}_{\mathrm{D}}[k]
\right\|_{F}^{2},
\label{precoding_objective}
\\
\mathrm{s.t.}\quad&
\left|
[\bm{F}_{\mathrm{RF}}]_{m,n}
\right|
=
\frac{1}{\sqrt{N_t}},
\quad
\forall m,n,
\label{precoding_cm_constraint}
\\
&
\left\|
\bm{F}_{\mathrm{RF}}
\bm{F}_{\mathrm{D}}[k]
\right\|_F^2
=
N_s,
\quad
\forall k,
\label{precoding_power_constraint}
\end{align}
\end{subequations}
where $\|\cdot\|_F$ denotes the Frobenius norm.
Constraint~\eqref{precoding_cm_constraint} enforces the constant-modulus
requirement of the analog phase-shifter network, while
\eqref{precoding_power_constraint} normalizes the hybrid precoder to
satisfy the transmit-power constraint at each subcarrier.
At the receiver side, the analog combiner and digital combiners can be optimized similarly based on an equivalent objective function of the squared error between the composite combiner $\bm{W}_{\mathrm{RF}}\bm{W}_{\mathrm{D}}[k]$ and the optimal solution $\bm{W}_{\mathrm{opt}}[k]$ under the constant-modulus constraint on $\bm{W}_{\mathrm{RF}}$.
The optimization can be obtained through a SOMP algorithm~\cite{lee2015hybrid}, which incurs considerable computational complexity. To speed up the optimization procedure in the online implementation, we propose a supervised-learning-based scheme where the labels of the analog precoder or analog combiner are generated through the SOMP algorithm to ensure the resultant performance. Specifically,
the analog precoder and analog combiner are obtained separately over predefined analog codebooks based on the SOMP algorithm, and the resultant indices of the optimal codewords serve as supervised labels for the proposed learning model. The digital precoders and digital combiners are then obtained based on the SVD of the equivalent channel response, which will be discussed in Section~\ref{Sec:Precoding}.


\section{Proposed Cascade XGBoost Hybrid Precoding}
\label{Sec:Precoding}

The proposed XGB-SVD framework separates analog beam prediction from
low-dimensional digital processing. As illustrated in
Fig.~\ref{fig_framework}, physical path parameters are first converted
into a fixed-dimensional feature vector and normalized using
feature-specific scaling. Cascade XGBoost then predicts the analog
transmit and receive beams, and SVD is finally applied to the
reduced-dimensional equivalent channel.

\subsection{SOMP Label Generation}

As shown in the offline-training part of Fig.~\ref{fig_framework}, SOMP generates the analog beam-index labels used to train the cascade classifiers.
Analog beam selection is formulated as a discrete classification
problem over finite beamforming codebooks.
Let
$\mathcal{F}=\{\bm{f}_1,\ldots,\bm{f}_{N_F}\}$
and
$\mathcal{W}=\{\bm{w}_1,\ldots,\bm{w}_{N_W}\}$
denote the transmit and receive analog beamforming codebooks,
respectively, where
$\bm{f}_{i}\in\mathbb{C}^{N_t\times1}$ and
$\bm{w}_{j}\in\mathbb{C}^{N_r\times1}$.
Here, $N_F$ and $N_W$ denote the corresponding codebook sizes.

The transmit and receive analog beamforming codebooks are constructed
using two-dimensional (2D) UPA discrete Fourier transform (DFT)
codebooks~\cite{ayach2014spatial} with angular oversampling to provide
finer beam resolution. Fourfold oversampling is applied along the azimuth dimension
and twofold oversampling along the elevation dimension, yielding
$N_F=512$ transmit codewords and $N_W=288$ receive codewords.

For the considered architecture,
$N_t^{\mathrm{RF}}=N_r^{\mathrm{RF}}\triangleq N_{\mathrm{RF}}$.
For each training realization, SOMP sequentially selects
$N_{\mathrm{RF}}$ transmit and $N_{\mathrm{RF}}$ receive codewords.
Let
$q_i^t\in\{1,\ldots,N_F\}$ and
$q_i^r\in\{1,\ldots,N_W\}$
denote the transmit and receive codeword indices selected at the SOMP step
$i\in\{1,\ldots,N_{\mathrm{RF}}\}$.
These indices are used as supervised labels for the cascade classifiers.


\subsection{Cascade XGBoost Beam Selection}

As shown in the Cascade XGBoost beam-selection block of
Fig.~\ref{fig_framework}, the cascade is designed to capture the
sequential dependence among beam selections and the correlation
between the transmit and receive decisions.
In SOMP, each newly selected beam depends on the
residual after the previously selected beams are removed. Moreover,
AoAs and AoDs associated with the same propagation paths induce
correlation between the transmit and receive beam indices.

Let $\mathcal{G}_{i}(\cdot)$ denote the pair of transmit-side and
receive-side XGBoost classifiers at cascade stage $i$. At the first
stage, the transmit--receive beam pair is predicted directly from the
physical feature vector as
$(\hat q_1^t,\hat q_1^r)=\mathcal{G}_1(\bm{p})$.

For $i>1$, the input to cascade stage $i$ is formed by augmenting the
physical feature vector with the normalized beam indices from the
preceding stages:
\begin{equation}
\bm{z}_i=
[\bm{p}^{T},
\bar q_1^t,\bar q_1^r,\ldots,
\bar q_{i-1}^t,\bar q_{i-1}^r]^{T},
\label{eq_cascade_input}
\end{equation}
where
$\bar q_j^t=(q_j^t-1)/N_F$ and
$\bar q_j^r=(q_j^r-1)/N_W$
denote the normalized transmit and receive codeword indices,
respectively. The next beam pair is predicted as
$(\hat q_i^t,\hat q_i^r)=\mathcal{G}_i(\bm{z}_i)$,
$i=2,\ldots,N_{\mathrm{RF}}$.

During training, the historical indices in $\bm{z}_i$ are the
ground-truth SOMP indices, corresponding to teacher forcing. During
online inference, they are replaced by the indices predicted at the
preceding cascade stages, resulting in an autoregressive prediction
process.

After all $N_{\mathrm{RF}}$ stages, the predicted analog beamformers
are constructed as
\begin{align}
\hat{\bm{F}}_{\mathrm{RF}}
&=
[\bm{f}_{\hat q_1^t},\ldots,
\bm{f}_{\hat q_{N_{\mathrm{RF}}}^t}],
\hat{\bm{W}}_{\mathrm{RF}}
&=
[\bm{w}_{\hat q_1^r},\ldots,
\bm{w}_{\hat q_{N_{\mathrm{RF}}}^r}].
\end{align}

The cascade therefore approximates the sequential beam-selection
structure of SOMP without executing the sparse reconstruction procedure
during online analog beam prediction.


\subsection{SVD-Based Digital Processing}

As shown in the online-inference part of Fig.~\ref{fig_framework}, SVD-based digital processing is applied after the analog beamformers have been predicted.
The proposed XGB-SVD architecture applies machine learning only to the analog beam-selection stage, while the digital beamformers are obtained analytically after RF-domain dimensionality reduction.
In contrast, the 
Fully XGB baseline uses the same physical path-parameter input and
predicts both the analog codeword indices and the digital beamforming
coefficients using XGBoost. Its analog stage follows the same cascade
beam-selection structure, whereas the digital stage is formulated as a
regression problem for the complex-valued baseband coefficients.
Prediction errors in these continuous-valued coefficients can therefore
lead to additional spectral-efficiency loss.

Given the predicted analog beamformers
$\hat{\bm{F}}_{\mathrm{RF}}$ and $\hat{\bm{W}}_{\mathrm{RF}}$, the
reduced-dimensional equivalent channel at subcarrier $k$ is
\begin{equation}
\bm{H}_{\mathrm{eq}}[k]
=
\hat{\bm{W}}_{\mathrm{RF}}^{H}
\bm{H}[k]
\hat{\bm{F}}_{\mathrm{RF}}.
\label{eq_equivalent_channel}
\end{equation}
Its singular value decomposition is
$\bm{H}_{\mathrm{eq}}[k]
=
\bm{U}_{k}\bm{\Sigma}_{k}\bm{V}_{k}^{H}$.
The digital precoder and combiner are then obtained from the dominant
$N_s$ singular vectors as
\begin{equation}
\bm{F}_{\mathrm{D}}[k]
=
\bm{V}_{k}(:,1:N_s),
\qquad
\bm{W}_{\mathrm{D}}[k]
=
\bm{U}_{k}(:,1:N_s).
\label{eq_digital_beamformers}
\end{equation}
Finally, $\bm{F}_{\mathrm{D}}[k]$ is normalized such that
$\|\hat{\bm{F}}_{\mathrm{RF}}\bm{F}_{\mathrm{D}}[k]\|_F^2=N_s$,
consistent with the transmit-power constraint in
\eqref{precoding_power_constraint}.


\section{Simulation Results}

\subsection{Simulation Setup}

The simulations evaluate both same-scenario performance and
cross-scenario transfer using physically distinct DeepMIMO ray-tracing
environments~\cite{alkhateeb2019deepmimo}. Outdoor1, Arizona State
University (ASU) Campus, and New York City (NYC) are used as sub-6-GHz
environments, whereas Houston City and Indoor represent mmWave
environments.

The principal system parameters are summarized in
Table~\ref{tab_parameters}.

\begin{table}[t]
\caption{Simulation Parameters}
\centering
\resizebox{1\columnwidth}{!}{
\begin{tabular}{lc}
\toprule
\textbf{Parameter} & \textbf{Value} \\
\midrule
Transmit antennas, $N_t$ & 64 \\
Receive antennas, $N_r$ & 36 \\
RF chains, $N_t^{\mathrm{RF}},N_r^{\mathrm{RF}}$ & 5 \\
Spatial streams, $N_s$ & 1, 2, 3 \\
OFDM subcarriers, $K$ & 128 (sub-6 GHz), 64 (mmWave) \\
Maximum retained paths, $L_{\max}$ & 10 \\
Transmit codebook size, $N_F$ & 512 \\
Receive codebook size, $N_W$ & 288 \\
Training samples & 10,000 \\
Validation samples & 10\% of training set \\
Testing samples & 1,000 \\
\bottomrule
\end{tabular}
}
\label{tab_parameters}
\end{table}

The considered benchmarks are summarized as follows.
1) \textbf{DNN-RT}: This baseline uses the same 70-dimensional
ray-tracing parameter vector as XGB-SVD and employs a DNN to predict
the analog beamforming codeword indices, followed by the same SVD-based
digital processing used in XGB-SVD~\cite{elbir2019cnn,li2019deep,mir2019machine}.
2) \textbf{DNN-H}: This baseline instead uses the reconstructed full
channel matrix as the DNN input and applies the same SVD-based digital
stage.
3) \textbf{Fully XGB}: This baseline uses the same physical
path-parameter input as XGB-SVD, but predicts both the analog codeword
indices and the complex-valued digital beamforming coefficients using
XGBoost.
4) \textbf{Fully Digital}: This benchmark employs
unconstrained fully digital water-filling over
the channel singular modes and serves as an
upper performance reference.
5) \textbf{MO-AltMin}: This method alternates between analog and digital
precoder optimization under the constant-modulus constraint
~\cite{yu2016alternating}.
6) \textbf{SOMP}: This method sequentially selects analog beamforming
codewords through sparse approximation of the fully digital precoder
~\cite{lee2015hybrid}.
7) \textbf{Random Phase}: This baseline uses randomly generated analog
beamforming phases without channel-adaptive beam selection.
8) \textbf{Random Phase+SVD}: This baseline uses random analog
beamformers followed by SVD-based digital processing on the resulting
equivalent channel.
XGBoost hyperparameters are optimized independently for each environment
using Optuna~\cite{akiba2019optuna}.

\begin{table}[t]
\caption{Characteristics of the DeepMIMO Environments}
\centering
\footnotesize
\renewcommand{\arraystretch}{0.92}
\resizebox{0.9\columnwidth}{!}{
\begin{tabular}{lccc}
\toprule
\textbf{Scenario} &
\textbf{Band} &
\textbf{Distance (m)} &
\textbf{Avg. Paths} \\
\midrule
Outdoor1     & Sub-6 GHz & 282--296 & 7.0 \\
ASU Campus   & Sub-6 GHz & 235--482 & 9.6 \\
NYC          & Sub-6 GHz & 4--393   & 8.7 \\
Houston City & mmWave    & 36--279  & 8.8 \\
Indoor       & mmWave    & 8--22    & 10.0 \\
\bottomrule
\end{tabular}
}
\label{tab_scenarios}
\end{table}

The five ray-tracing environments exhibit different propagation
geometries, as summarized in Table~\ref{tab_scenarios}.
Outdoor1 has a narrow user-distance range, whereas ASU Campus
and NYC span broader ranges, with NYC covering the widest
interval among the sub-6-GHz scenarios.
Indoor represents a short-range mmWave setting, with an
average number of paths reaching $L_{\max}=10$.


\subsection{Same-Scenario Performance}

\begin{figure}[t]
\centering
\includegraphics[width=\columnwidth]
{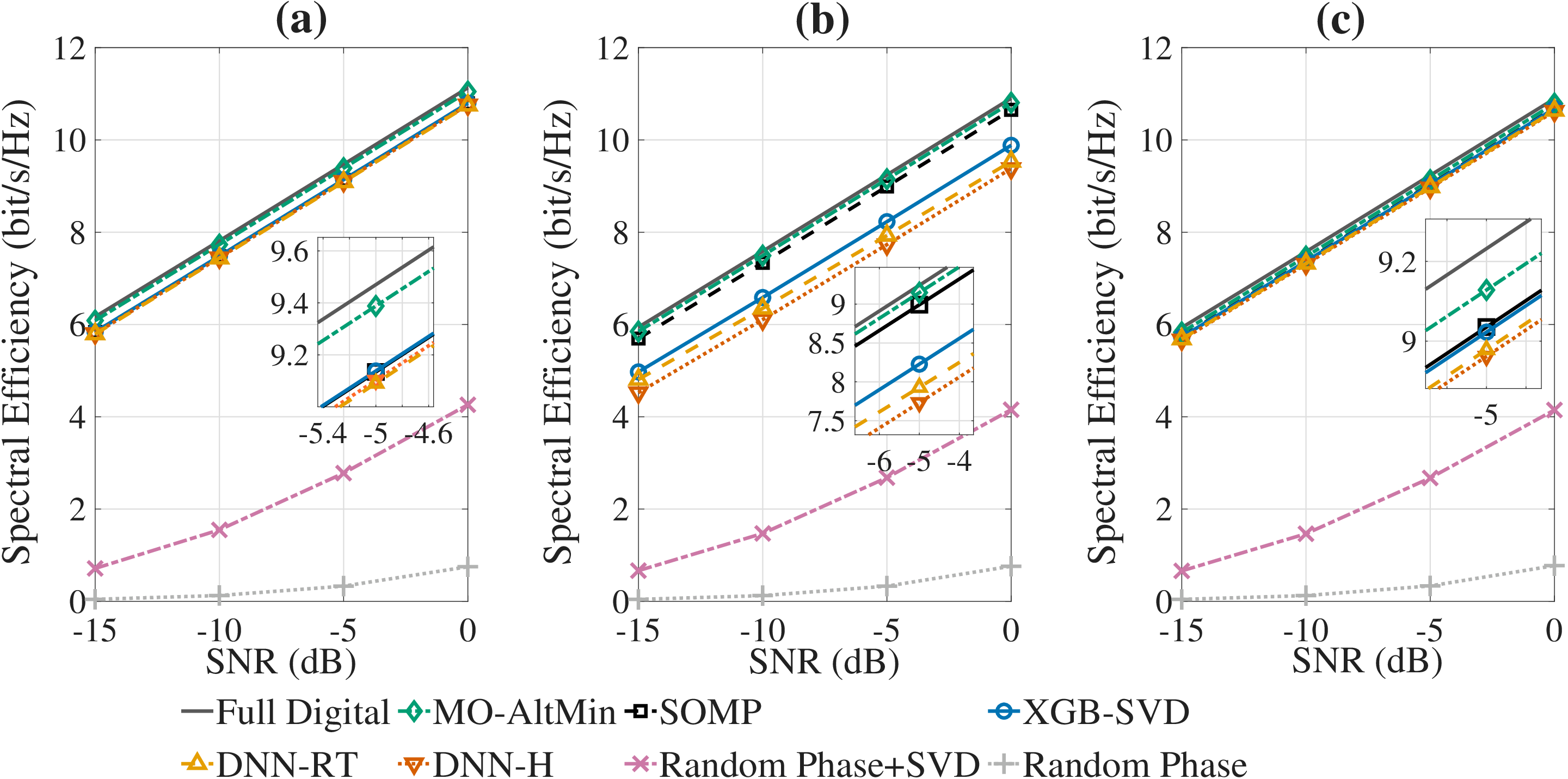}
\caption{Spectral efficiency in representative DeepMIMO environments: (a) Outdoor1, (b) NYC, and (c) Indoor.}
\label{fig_main_results}
\end{figure}

Fig.~\ref{fig_main_results} compares the same-scenario spectral
efficiency in (a) Outdoor1, (b) NYC, and (c) Indoor. XGB-SVD
consistently outperforms the considered DNN baselines across these
representative environments.

In Outdoor1, XGB-SVD closely approaches SOMP and MO-AltMin over the
evaluated SNR range. A similarly small gap is observed in Indoor, where
XGB-SVD remains close to SOMP and the fully digital reference. In
contrast, the performance gap becomes more pronounced in NYC,
indicating that the mapping from physical path parameters to analog
codeword combinations is more difficult in this environment.

These results suggest that the difficulty of analog beam selection
depends strongly on the propagation environment rather than on the
carrier-frequency category alone. Indoor provides one of the strongest
results despite operating at mmWave frequencies, whereas NYC remains
substantially more challenging in the sub-6-GHz band.


\subsection{Ablation Study}

\begin{figure}[t]
\centering
\includegraphics[width=\linewidth]
{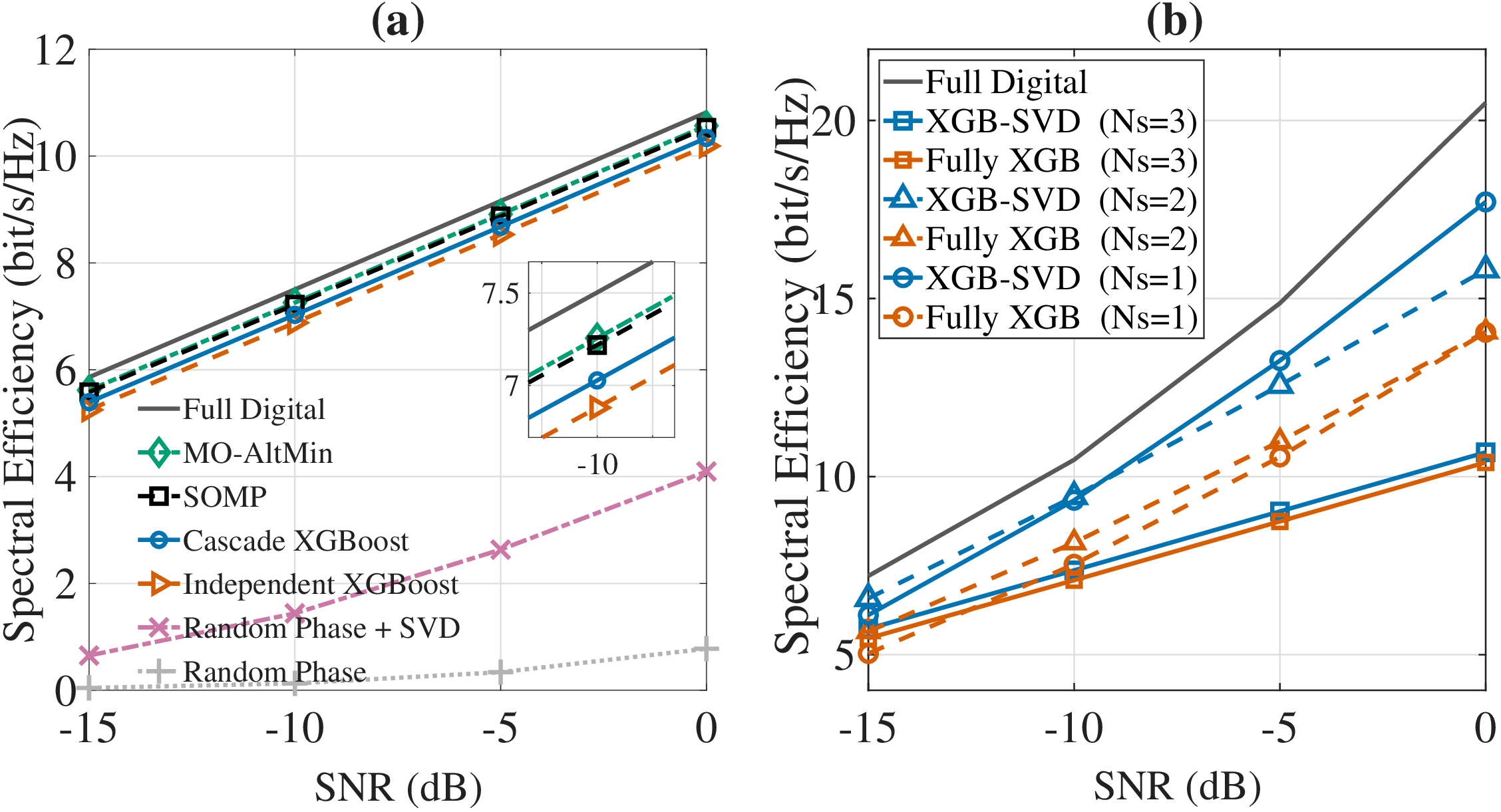}
\caption{Ablation study of the proposed framework: (a) independent XGBoost versus Cascade XGBoost and (b) Fully XGB versus XGB-SVD.}
\label{fig_ablation}
\end{figure}

Fig.~\ref{fig_ablation} presents two ablation studies:
(a) independent XGBoost versus Cascade XGBoost and
(b) Fully XGB versus XGB-SVD. The results evaluate the contributions
of the cascade structure and the SVD-based digital stage, respectively.

Cascade XGBoost outperforms independent XGBoost because previous beam
decisions provide useful information for subsequent classifications.
This behavior is consistent with the sequential residual update of
SOMP and with the physical correlation between AoAs and AoDs
associated with the same propagation paths.

XGB-SVD also outperforms Fully XGB because digital beamforming is more
sensitive to continuous-valued regression errors.
Analog beam selection is a discrete classification problem over finite
codebooks, whereas digital precoding requires accurate complex-valued
coefficients. Retaining SVD over the reduced-dimensional equivalent
channel therefore yields a better performance-complexity trade-off.


\subsection{Cross-Scenario Transfer}
\vspace{-0.1in}

\begin{figure}[t]
\centering
\includegraphics[width=\columnwidth]
{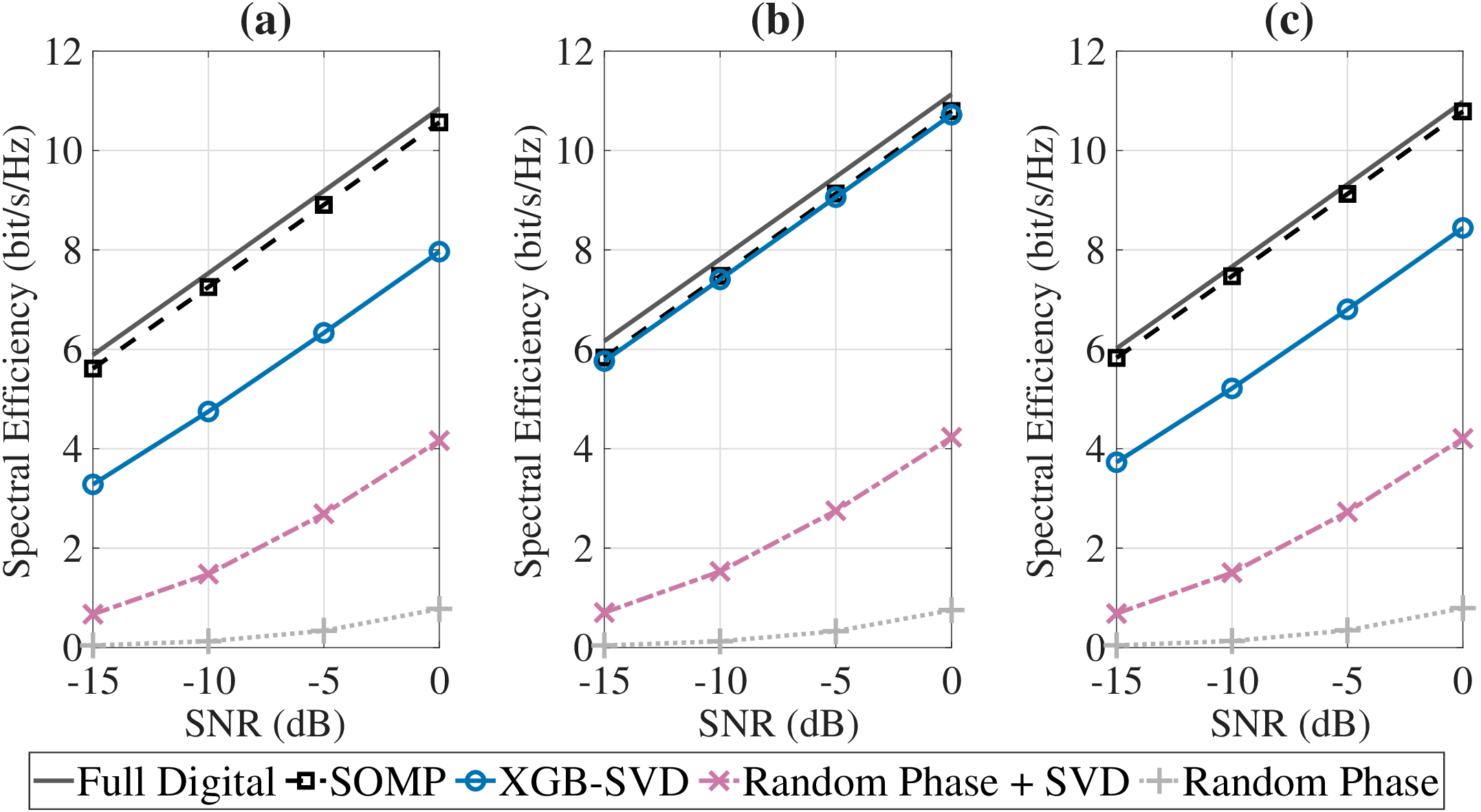}
\caption{Representative cross-scenario transfer performance of XGB-SVD: (a) Outdoor1 to ASU Campus, (b) ASU Campus to Outdoor1, and (c) NYC to Boston City.}
\label{fig_cross}
\end{figure}

Fig.~\ref{fig_cross} evaluates cross-scenario transfer
without retraining, including Outdoor1-to-ASU Campus,
ASU Campus-to-Outdoor1, and NYC-to-Boston City.
Boston City is excluded entirely from training.
The results reveal asymmetric transfer behavior:
training on ASU Campus generalizes more effectively
to Outdoor1 than the reverse direction.

This asymmetry suggests that training-environment
diversity plays an important role in generalization.
Models trained in propagation-diverse environments,
such as ASU Campus and NYC, exhibit stronger transfer
performance, including in the unseen Boston City
scenario. These observations highlight the potential
benefit of diverse training environments for
cross-scenario hybrid precoding.


\subsection{Multi-Stream Transmission}

Fig.~\ref{fig_ablation}(b) highlights the comparison between XGB-SVD
and Fully XGB for $N_s=1$, 2, and 3 spatial streams. Increasing the
number of streams provides higher spatial multiplexing gain while
imposing a more demanding analog beam-selection task.

XGB-SVD remains close to SOMP in Outdoor1 and Indoor as the number of
streams increases, whereas larger performance gaps are observed in ASU
Campus and Houston City. NYC exhibits the largest degradation among the
considered environments as $N_s$ increases.

At $N_s=3$, the difference between XGB-SVD and SOMP reaches
approximately 3 bit/s/Hz at high SNR in NYC. These results suggest that
the difficulty of analog codeword prediction increases when spatial
multiplexing is combined with more challenging propagation conditions.


\subsection{Computational Complexity}

The proposed XGB-SVD architecture reduces the arithmetic burden
associated with neural hybrid precoding while retaining a
low-dimensional model-based digital stage.
We quantify arithmetic complexity using floating-point operations
(FLOPs) and tree-inference complexity using XGBoost node comparisons.
Table~\ref{tab_complexity} summarizes the online computational
requirements and offline training time under the Outdoor1
configuration.

\begin{table}[t]
\caption{Computational Complexity and Training Time}
\centering
\footnotesize
\renewcommand{\arraystretch}{0.92}
\setlength{\tabcolsep}{2pt}
\resizebox{0.9\columnwidth}{!}{
\begin{tabular}{lccc}
\toprule
\textbf{Method} &
\textbf{FLOPs} &
\textbf{XGBoost Comparisons} &
\textbf{Training Time} \\
\midrule
SOMP      & 2.08 M  & 0      & -- \\
Fully XGB & 0       & 0.67 M & 480.2 min \\
XGB-SVD   & 0.67 M  & 0.67 M & 35.0 min \\
DNN-RT    & 4.35 M  & 0      & 49.2 min \\
DNN-H     & 14.83 M & 0      & 61.7 min \\
\bottomrule
\end{tabular}
}
\label{tab_complexity}
\end{table}

The operation counts in Table~\ref{tab_complexity} represent different
types of computation and should not be directly equated.
FLOPs denote arithmetic operations, while XGBoost comparisons primarily
correspond to tree-node threshold comparisons.

XGB-SVD requires a limited number of arithmetic operations after
analog dimension reduction.
It uses approximately 0.67 million XGBoost comparisons for analog beam
prediction and 0.67 million FLOPs for the digital stage. In contrast,
DNN-RT and DNN-H require approximately 4.35 and 14.83 million FLOPs,
respectively.

The proposed method also provides favorable offline training
efficiency.
Under the same central processing unit (CPU) environment, XGB-SVD
requires 35 min of training time,
which is approximately 29\% and 43\% lower than DNN-RT and DNN-H,
respectively. Fully XGB requires 480.2 min because the
high-dimensional continuous digital coefficients must also be learned.
Replacing this regression stage with reduced-dimensional SVD reduces
the training time by approximately 92.7\%.

Overall, the complexity results support the proposed division of
computation between learning and signal processing.
Low-dimensional physical features reduce the learning input size,
tree-based inference avoids the dense arithmetic operations associated
with DNNs, and the digital beamforming stage is solved analytically
after RF-domain dimensionality reduction.

\vspace{-0.1in}
\section{Conclusion}

This paper proposed XGB-SVD, a lightweight hybrid precoding framework
that maps compact ray-tracing parameters to analog beam codewords using
Cascade XGBoost and obtains digital beamformers from the
reduced-dimensional equivalent channel via SVD. Across multiple
DeepMIMO environments, XGB-SVD outperforms the considered DNN baselines,
approaches SOMP in several scenarios, and provides lower computational
and training costs. Cross-scenario results further indicate the benefit
of diverse training environments. Future work will consider imperfect
path estimates and multi-environment training.


\end{document}